\documentclass[onecolumn]{article}
\usepackage{a4wide}
\usepackage{amsmath}
\usepackage{epsfig}

\newcommand{\define}{\stackrel{\Delta}{=}}

\newcommand{\PAMv}{{\cal{A}}_{\text{4-PAM}}}

\newcommand{\QAMv}{{\cal{A}}_{\text{4-QAM}}}
\newcommand{\QAMz}{{\cal{A}}_{\text{16-QAM}}}

\newcommand{\uy}{\underline{y}}
\newcommand{\ua}{\underline{a}}
\newcommand{\ub}{\underline{b}}
\newcommand{\uc}{\underline{c}}
\newcommand{\ud}{\underline{d}}
\newcommand{\ue}{\underline{e}}
\newcommand{\un}{\underline{n}}
\newcommand{\uz}{\underline{z}}

\newcommand{\ux}{\underline{x}}
\newcommand{\SNR}{\text{SNR}}

\begin{document}
\title{Rotated and Scaled Alamouti Coding}
\author{Frans M.J. Willems\thanks{Philips Research Laboratories, High Tech Campus 37, 5656AE Eindhoven, The Netherlands}}

\maketitle
\begin{abstract}
Repetition-based retransmission is used in Alamouti-modulation [1998] for $2\times 2$ MIMO systems.
We propose to use instead of ordinary repetition so-called "scaled repetition" together with  rotation.
It is shown that the rotated and scaled Alamouti code has a hard-decision performance which is only slightly worse than that of the Golden code [2005], the best known $2\times 2$ space-time code.
Decoding the Golden code requires an exhaustive search over all codewords, while our rotated and scaled Alamouti code can be decoded with an acceptable complexity however.
\end{abstract}
\section{Scaled-repetition Retransmission for the SISO Channel}\label{sec:scarepsiso}
First we consider transmission over a single-input single-output (SISO) additive white Gaussian noise (AWGN) channel (see Fig. \ref{fig:awgn}), and introduce scaled-repetition retransmission.
It turns out that scaled-repetition improves upon ordinary-repetition retransmission.
\subsection{Some information theory}
\begin{figure}[h]
\centering
\setlength{\unitlength}{3947sp}%
\begingroup\makeatletter\ifx\SetFigFont\undefined%
\gdef\SetFigFont#1#2#3#4#5{%
  \reset@font\fontsize{#1}{#2pt}%
  \fontfamily{#3}\fontseries{#4}\fontshape{#5}%
  \selectfont}%
\fi\endgroup%
\begin{picture}(3174,988)(214,1052)
\thinlines
\put(1801,1289){\oval(450,450)}
\put(2476,1064){\framebox(900,450){}}
\put(226,1064){\framebox(900,450){}}
\put(2026,1289){\vector( 1, 0){450}}
\put(1126,1289){\vector( 1, 0){450}}
\put(1801,1964){\vector( 0,-1){450}}
\put(2926,1233){\makebox(0,0)[b]{\smash{{\SetFigFont{10}{12.0}{\rmdefault}{\mddefault}{\updefault}receiver}}}}
\put(676,1233){\makebox(0,0)[b]{\smash{{\SetFigFont{10}{12.0}{\rmdefault}{\mddefault}{\updefault}transmitter}}}}
\put(1913,1851){\makebox(0,0)[lb]{\smash{{\SetFigFont{10}{12.0}{\rmdefault}{\mddefault}{\updefault}$n$}}}}
\put(2251,1514){\makebox(0,0)[b]{\smash{{\SetFigFont{10}{12.0}{\rmdefault}{\mddefault}{\updefault}$y$}}}}
\put(1351,1514){\makebox(0,0)[b]{\smash{{\SetFigFont{10}{12.0}{\rmdefault}{\mddefault}{\updefault}$x$}}}}
\put(1801,1233){\makebox(0,0)[b]{\smash{{\SetFigFont{12}{14.4}{\rmdefault}{\mddefault}{\updefault}$+$}}}}
\end{picture}
\caption{The AWGN channel.}
\label{fig:awgn}
\end{figure}
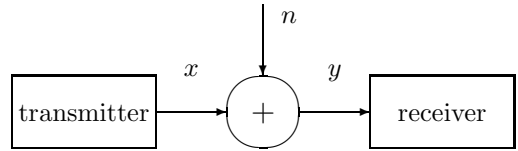
The real-valued output $y_k$ for transmission $k=1,2,\cdots,K,$ see Fig. \ref{fig:awgn}, satisfies
\begin{equation}
y_k=x_k+n_k,
\end{equation}
where $x_k$ is the real-valued channel input for transmission $k$ and $n_k$ is a real-valued Gaussian noise sample with mean $E[N_k]=0,$ variance $E[N_k^2]=\sigma^2$, which is uncorrelated with all other noise samples.
The transmitter power is limited, i.e. we require that $E[X_k^2]\leq P$.
It is well-known that an $X$ which is Gaussian with mean 0 and variance $P$ achieves capacity.
This basic capacity (in bit/transm.) equals
\begin{equation}
C= \frac12\log_2(1 + \frac{P}{\sigma^2}).
\end{equation}
When we retransmit (repeat) codewords , each symbol $x_k$ from such a codeword $(x_1,x_2,\cdots,x_K)$ is actually transmitted and received twice, i.e. $x_{k1}=x_{k2}=x_k$, and
\begin{equation}
y_{k1} = x_{k} + n_{k1}, \text{ and }
y_{k2} = x_{k} + n_{k2}. \label{eq:ordinaryrep}
\end{equation}
An optimal receiver can form $z_{k} = \frac{y_{k1}+y_{k2}}{2} = x_k + \frac{n_{k1}+n_{k2}}{2}$.
Now the variance of the noise variable $(N_{k1}+N_{k2})/2$ is $\sigma^2/2$.
Therefore the repetition capacity for a single repetition in bit/transm. is
\begin{equation}
C_{r} = \frac14 \log_2(1+\frac{2P}{\sigma^2}).
\end{equation}
Fig. \ref{plot:sisocaps} shows the basic capacity $C$ (black line) and repetition capacity $C_{r}$ (blue line) as a function of the signal-to-noise ratio $\SNR$ which is defined as
\begin{equation}
\SNR \define  P/\sigma^2.
\end{equation}
It is easy to see that always $C_{r} \leq C$.
For large $\SNR$ we may write $C_r \approx C/2 +1/4$, while for small $\SNR$ we obtain $C_r \approx C$.
\subsection{Ordinary and scaled repetition for 4-PAM}
\begin{figure}
\centering
\setlength{\unitlength}{3158sp}%
\begingroup\makeatletter\ifx\SetFigFont\undefined%
\gdef\SetFigFont#1#2#3#4#5{%
  \reset@font\fontsize{#1}{#2pt}%
  \fontfamily{#3}\fontseries{#4}\fontshape{#5}%
  \selectfont}%
\fi\endgroup%
\begin{picture}(3893,1968)(326,-3560)
\thinlines
\put(3827,-2535){\oval(112,112)}
\put(3377,-3435){\oval(112,112)}
\put(2927,-2085){\oval(112,112)}
\put(2477,-2985){\oval(112,112)}
\put(1802,-2085){\oval(112,112)}
\put(1352,-2535){\oval(112,112)}
\put(902,-2985){\oval(112,112)}
\put(452,-3435){\oval(112,112)}
\put(3151,-3548){\line( 0, 1){1800}}
\put(2363,-2761){\line( 1, 0){1800}}
\put(1126,-3548){\line( 0, 1){1800}}
\put(338,-2761){\line( 1, 0){1800}}
\put(2476,-2086){\line( 1, 0){1350}}
\put(2476,-2536){\line( 1, 0){1350}}
\put(2476,-2986){\line( 1, 0){1350}}
\put(3826,-2086){\line( 0,-1){1350}}
\put(3376,-2086){\line( 0,-1){1350}}
\put(2926,-2086){\line( 0,-1){1350}}
\put(2476,-2086){\line( 0,-1){1350}}
\put(2476,-3436){\line( 1, 0){1350}}
\put(1351,-2086){\line( 0,-1){1350}}
\put(901,-2086){\line( 0,-1){1350}}
\put(451,-3436){\line( 1, 0){1350}}
\put(451,-2986){\line( 1, 0){1350}}
\put(451,-2536){\line( 1, 0){1350}}
\put(1801,-2086){\line( 0,-1){1350}}
\put(451,-2086){\line( 0,-1){1350}}
\put(451,-2086){\line( 1, 0){1350}}
\put(3854,-2873){\makebox(0,0)[lb]{\smash{{\SetFigFont{6}{7.2}{\rmdefault}{\mddefault}{\updefault}+3}}}}
\put(1829,-2873){\makebox(0,0)[lb]{\smash{{\SetFigFont{6}{7.2}{\rmdefault}{\mddefault}{\updefault}+3}}}}
\put(3404,-2873){\makebox(0,0)[lb]{\smash{{\SetFigFont{6}{7.2}{\rmdefault}{\mddefault}{\updefault}+1}}}}
\put(1379,-2873){\makebox(0,0)[lb]{\smash{{\SetFigFont{6}{7.2}{\rmdefault}{\mddefault}{\updefault}+1}}}}
\put(2955,-2873){\makebox(0,0)[lb]{\smash{{\SetFigFont{6}{7.2}{\rmdefault}{\mddefault}{\updefault}-1}}}}
\put(930,-2873){\makebox(0,0)[lb]{\smash{{\SetFigFont{6}{7.2}{\rmdefault}{\mddefault}{\updefault}-1}}}}
\put(2505,-2873){\makebox(0,0)[lb]{\smash{{\SetFigFont{6}{7.2}{\rmdefault}{\mddefault}{\updefault}-3}}}}
\put(480,-2873){\makebox(0,0)[lb]{\smash{{\SetFigFont{6}{7.2}{\rmdefault}{\mddefault}{\updefault}-3}}}}
\put(3178,-3407){\makebox(0,0)[lb]{\smash{{\SetFigFont{6}{7.2}{\rmdefault}{\mddefault}{\updefault}-3}}}}
\put(3179,-2957){\makebox(0,0)[lb]{\smash{{\SetFigFont{6}{7.2}{\rmdefault}{\mddefault}{\updefault}-1}}}}
\put(3179,-2508){\makebox(0,0)[lb]{\smash{{\SetFigFont{6}{7.2}{\rmdefault}{\mddefault}{\updefault}+1}}}}
\put(3178,-2059){\makebox(0,0)[lb]{\smash{{\SetFigFont{6}{7.2}{\rmdefault}{\mddefault}{\updefault}+3}}}}
\put(1153,-3407){\makebox(0,0)[lb]{\smash{{\SetFigFont{6}{7.2}{\rmdefault}{\mddefault}{\updefault}-3}}}}
\put(3207,-1862){\makebox(0,0)[lb]{\smash{{\SetFigFont{10}{12.0}{\rmdefault}{\mddefault}{\updefault}$x_{k2}$}}}}
\put(1182,-1861){\makebox(0,0)[lb]{\smash{{\SetFigFont{10}{12.0}{\rmdefault}{\mddefault}{\updefault}$x_{k2}$}}}}
\put(4219,-2649){\makebox(0,0)[rb]{\smash{{\SetFigFont{10}{12.0}{\rmdefault}{\mddefault}{\updefault}$x_{k1}$}}}}
\put(2194,-2649){\makebox(0,0)[rb]{\smash{{\SetFigFont{10}{12.0}{\rmdefault}{\mddefault}{\updefault}$x_{k1}$}}}}
\put(1153,-2059){\makebox(0,0)[lb]{\smash{{\SetFigFont{6}{7.2}{\rmdefault}{\mddefault}{\updefault}+3}}}}
\put(1154,-2957){\makebox(0,0)[lb]{\smash{{\SetFigFont{6}{7.2}{\rmdefault}{\mddefault}{\updefault}-1}}}}
\put(1154,-2508){\makebox(0,0)[lb]{\smash{{\SetFigFont{6}{7.2}{\rmdefault}{\mddefault}{\updefault}+1}}}}
\end{picture}%
\caption{Two mappings from  $x_{k1}$ to $x_{k2}$. On the right the
scaled-repetition mapping, left the ordinary-repetition mapping.}
\label{fig:permmodul}
\end{figure}
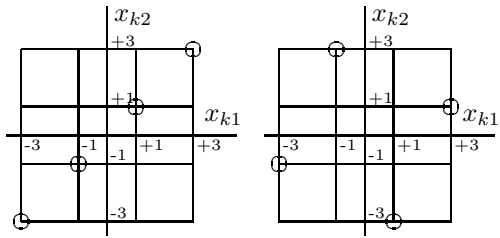
\begin{figure}
\centering
\epsfig{file=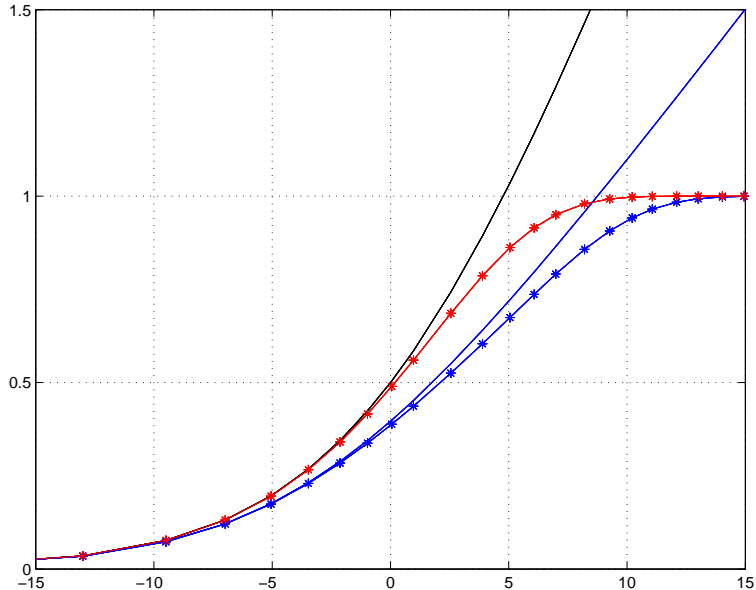,width=100mm}
\caption{Basic capacity $C$ (black curve) and repetition capacity $C_{r}$ (blue) in bit/transm. as a function of $\SNR=P/\sigma^2$ in dB (horizontally).
Also the maximum transmission rates achievable with 4-PAM in the ordinary-repetition case (blue *'s).
In red *'s the maximum rates achievable using scaled-repetition mapping.}
\label{plot:sisocaps}
\end{figure}

When we use 4-PAM modulation, the channel inputs $x_k$ assume values from $\PAMv =\{ -3,-1, +1,$ $+3 \}$,
each with probability $1/4$.
Ordinary repetition, see (\ref{eq:ordinaryrep}), leads to signal points $(x_1,x_2) = (x,x)$ for $x\in\PAMv$, see the left part of Fig. \ref{fig:permmodul}.
For this case the maximum transmission rate $I(X;Y_1,Y_2)$ is shown in Fig. \ref{plot:sisocaps} with blue asterisks.
Note that this maximum transmission rate is slightly smaller than the corresponding capacities $C_r,$ mainly because uniform inputs are used instead of Gaussians.

We can use Benelli's \cite{Benn1987} method to improve upon ordinary-repetition retransmission, i.e. by modulating the retransmitted symbol differently. We could e.g. take
\begin{equation} x_{k1} = x_{k}, \text{ and } x_{k2} = M_2( x_{k}) \text{ for } x_{k}\in\PAMv,
\end{equation}
where $M_2(\alpha)=2\alpha -5$ if $\alpha>0$ and $M_2(\alpha)=2\alpha+5$ for $\alpha<0$.
We call this method {\em scaled repetition} since we scale a symbol by a factor (2 here) and then compensate (add -5 or +5) in order to obtain a symbol from $\PAMv$.
This results in the signal points $(x,M_2(x))$ for $x\in \PAMv$, see Fig. \ref{fig:permmodul}, right part.
Also for the scaled-repetition case the maximum transmission rate $I(X;Y_1,Y_2)$ is shown in figure \ref{plot:sisocaps}, now with red asterisks.
Note that this maximum transmission rate is only slightly smaller than the basic capacity $C.$
Ordinary repetition is however definitively inferior to the basic transmission if the $\SNR$ is not very small.
\subsection{Demodulation complexity}
Scaled repetition outperforms ordinary repetition, but also has a disadvantage.
In an ordinary-repetition system the output $y_{k}=(y_{k1}+y_{k2})/2$ is simply sliced.
In a system that uses scaled repetition we can only slice after having distinguished between two
cases.
More precisely note that $x_{k2} = M_2(x_{k}) = 2x_{k} - D_2(x_{k})$,
where $D_2(\alpha) = 5$ if $\alpha>0$ and $D_2(\alpha)=-5$ if $\alpha<0$.
Now we can use a slicer for $y_{k1}+2y_{k2} = x_k + n_{k1} + 2( 2x_k - D_2(x_k) + n_{k2}) = 5 x_k - 2D_2(x_k) +n_{k1}+2n_{k2}$.
Assuming that $x_k\in\{-3,-1\}$ we get that $D_2(x_k)=-5$ and this implies that we should put a threshold at 0 to distinguish between $-3$ and $-1$.
Similarly assuming that $x_k\in\{+1,+3\}$ we get $D_2(x_k)=5$ and we must slice $y_{k1}+2y_{k2}$ again with a threshold at 0.
Then the best overall candidate $\hat{x_k}$ is found by minimizing $(y_{k1}-\hat{x_k})^2 +(y_{k2}-M_2(\hat{x_k}) )^2$ over the two candidates.
\section{Fundamental Properties for the $2 \times 2$ MIMO Channel}
\subsection{Model description}
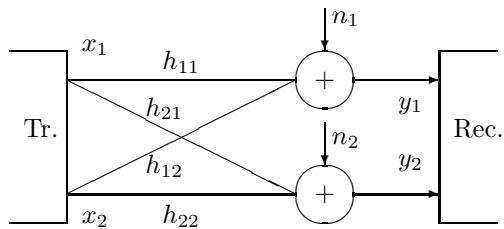
\begin{figure}[h]
\centering
\setlength{\unitlength}{3158sp}%
\begingroup\makeatletter\ifx\SetFigFont\undefined%
\gdef\SetFigFont#1#2#3#4#5{%
  \reset@font\fontsize{#1}{#2pt}%
  \fontfamily{#3}\fontseries{#4}\fontshape{#5}%
  \selectfont}%
\fi\endgroup%
\begin{picture}(3849,1923)(1114,-1056)
\thinlines
\put(3601,164){\oval(450,450)}
\put(3601,-736){\oval(450,450)}
\put(1126,389){\line( 1, 0){450}}
\put(1576,389){\line( 0,-1){1350}}
\put(1576,-961){\line(-1, 0){450}}
\put(1576,-736){\line( 2, 1){1800}}
\put(3376,164){\line(-1, 0){1800}}
\put(1576,164){\line( 2,-1){1800}}
\put(3376,-736){\line(-1, 0){1800}}
\put(4951,389){\line(-1, 0){450}}
\put(4501,389){\line( 0,-1){1350}}
\put(4501,-961){\line( 1, 0){450}}
\put(3826,164){\vector( 1, 0){675}}
\put(3826,-736){\vector( 1, 0){675}}
\put(3601,-173){\vector( 0,-1){338}}
\put(3601,727){\vector( 0,-1){338}}
\put(4613,-286){\makebox(0,0)[lb]{\smash{{\SetFigFont{10}{12.0}{\rmdefault}{\mddefault}{\updefault}Rec.}}}}
\put(1238,-286){\makebox(0,0)[lb]{\smash{{\SetFigFont{10}{12.0}{\rmdefault}{\mddefault}{\updefault}Tr.}}}}
\put(2476,249){\makebox(0,0)[b]{\smash{{\SetFigFont{10}{12.0}{\rmdefault}{\mddefault}{\updefault}$h_{11}$}}}}
\put(2475,-961){\makebox(0,0)[b]{\smash{{\SetFigFont{10}{12.0}{\rmdefault}{\mddefault}{\updefault}$h_{22}$}}}}
\put(2194,-117){\makebox(0,0)[lb]{\smash{{\SetFigFont{10}{12.0}{\rmdefault}{\mddefault}{\updefault}$h_{21}$}}}}
\put(2195,-567){\makebox(0,0)[lb]{\smash{{\SetFigFont{10}{12.0}{\rmdefault}{\mddefault}{\updefault}$h_{12}$}}}}
\put(1800,390){\makebox(0,0)[b]{\smash{{\SetFigFont{10}{12.0}{\rmdefault}{\mddefault}{\updefault}$x_1$}}}}
\put(1800,-960){\makebox(0,0)[b]{\smash{{\SetFigFont{10}{12.0}{\rmdefault}{\mddefault}{\updefault}$x_2$}}}}
\put(3601,108){\makebox(0,0)[b]{\smash{{\SetFigFont{10}{12.0}{\rmdefault}{\mddefault}{\updefault}$+$}}}}
\put(3601,-792){\makebox(0,0)[b]{\smash{{\SetFigFont{10}{12.0}{\rmdefault}{\mddefault}{\updefault}$+$}}}}
\put(4275,-60){\makebox(0,0)[b]{\smash{{\SetFigFont{10}{12.0}{\rmdefault}{\mddefault}{\updefault}$y_1$}}}}
\put(4275,-510){\makebox(0,0)[b]{\smash{{\SetFigFont{10}{12.0}{\rmdefault}{\mddefault}{\updefault}$y_2$}}}}
\put(3875,614){\makebox(0,0)[rb]{\smash{{\SetFigFont{10}{12.0}{\rmdefault}{\mddefault}{\updefault}$n_1$}}}}
\put(3875,-342){\makebox(0,0)[rb]{\smash{{\SetFigFont{10}{12.0}{\rmdefault}{\mddefault}{\updefault}$n_2$}}}}
\end{picture}%
\caption{Model of a $2 \times 2$ MIMO channel.}
\label{fig:mimo}
\vspace{-2mm}
\end{figure}
Next consider a $2 \times 2$ MIMO channel (see Fig. \ref{fig:mimo}).
Both the transmitter and the receiver use two antennas.
The output vector $(y_{1k},y_{2k})$ at transmission $k$ relates to the corresponding input vector  $(x_{1k},x_{2k})$ as given by
\begin{equation}
\left(\begin{array}{c}
   y_{1k} \\
   y_{2k} \end{array}\right)
=
\left(\begin{array}{cc}
        h_{11}        & h_{12}  \\
        h_{21}        & h_{22}
      \end{array} \right)
\left(\begin{array}{c} x_{1k} \\ x_{2k} \end{array} \right)
+
\left(\begin{array}{c}
   n_{1k} \\
   n_{2k}
    \end{array}\right)
\end{equation}
where $(N_{1k},N_{2k})$ is a pair of independent zero-mean circularly symmetric complex Gaussians, both having variance $\sigma^2$ (per two dimensions).
Noise variable pairs in different transmissions are independent.

We assume that the four channel coefficients $H_{11}, H_{12}, H_{21}$, and $H_{22}$ are independent zero-mean circularly symmetric complex Gaussians, each having variance 1 (per two dimensions).
{\em The channel coefficients are chosen prior to a block of $K$ transmissions and remain constant over that block.}

The complex transmitted symbols $(X_{k1},X_{k2})$ must satisfy a power constraint, i.e.
\begin{equation}
E[  X_{k1}X^{*}_{k1} + X_{k2}X^{*}_{k2} ] \leq P.
\end{equation}

\subsection{Telatar capacity}
If the channel input variables are independent zero-mean circularly symmetric complex Gaussians both having variance $P/2,$ then the resulting mutual information (called Telatar capacity here, see \cite{Tela1999}) is\footnote{Here $H^{\dagger}$ denotes the Hermitian transpose of $H$. It involves both transposition and complex conjugation.}
\begin{equation}
C_{\text{Telatar}}(H)= \log_2 \det (I_2 + \frac{P/2}{\sigma^2} H H^{\dagger} ),
\end{equation}
where $H = \left(\begin{array}{cc} h_{11} & h_{12} \\
                        h_{21} & h_{22}  \end{array}\right)$,
i.e. the actual channel-coefficient matrix and $I_2$ the $2\times 2$ identity matrix.
Also in the $2\times 2$ MIMO case we define the signal-to-niose ratio as
\begin{equation}
\SNR \define P/\sigma^2.
\end{equation}
It can be shown (see e.g. Yao (\cite{Yao2003}, p. 36) that for fixed $R$ and $\SNR$ large enough
$ \Pr\{C_{\text{Telatar}(H)} < R \} \approx \gamma \cdot \SNR^{-4},$ for some constant $\gamma$.

\subsection{Worst-case error-probabilities}
Consider $M$ (one for each message)   $K\times 2$ code-matrices $\uc_1,\uc_2,\cdots,\uc_M$ resulting in a unit average energy code.
Then Tarokh, Seshadri and Calderbank \cite{TaroSeshCald1998} showed that for large $\SNR$
\begin{equation}
\Pr\{\uc\rightarrow \uc'\} \approx \gamma' (\det( (\uc'-\uc)(\uc'-\uc)^{\dagger} ) ^{-2} \SNR^{-4}.
\end{equation}
for some $\gamma'$ if the rank of the difference matrices $\uc-\uc'$ is 2, and we transmit $\ux=\sqrt{P} \uc$.
If this holds for all difference matrices we say that the diversity order is 4.
Therefore it makes sense to maximize the minimum modulus of the determinant over all code-matrix differences.
\section{Alamouti: Ordinary Repetition}
Alamouti \cite{Alam1998} proposed a modulation scheme (space-time code) for the $2\times 2$ MIMO cannel which allows for a very simple detector.
Two complex symbols $s_1$ and $s_2$ are transmitted in the first transmission (an odd transmission) and in the second transmission (the next even transmission) these symbols are more or less repeated.
More precisely
\begin{equation} \label{eq:Alamouti}
\left(\begin{array}{cc} x_{11} & x_{12} \\
                         x_{21} & x_{22} \end{array}\right)
=
\left(\begin{array}{cc} s_{1} & -s^{*}_{2} \\
                        s_{2} &  s^{*}_{1} \end{array}\right).
\end{equation}
The received signal is now
\begin{equation}
\left(\begin{array}{cc} y_{11} & y_{12} \\
                        y_{21} & y_{22} \end{array}\right)
=
 \left(\begin{array}{cc} h_{11} & h_{12} \\
                        h_{21} & h_{22}  \end{array}\right)
 \left(\begin{array}{cc} s_{1} & -s^{*}_{2} \\
                            s_{2} & s^{*}_{1} \end{array}\right)
+
\left(\begin{array}{cc} n_{11} & n_{12} \\
                        n_{21} & n_{22} \end{array}\right).
\end{equation}
Rewriting this results in
\begin{equation}
\left(\begin{array}{c}
   y_{11} \\
   y_{21} \\
   y^{*}_{12}\\
   y^{*}_{22} \end{array}\right)
=
\left(\begin{array}{cc}
        h_{11}         & h_{12}  \\
        h_{21}        & h_{22} \\
        h^{*}_{12} & -h^{*}_{11} \\
        h^{*}_{22} & -h^{*}_{21} \\
      \end{array} \right)
\left(\begin{array}{c} s_1 \\ s_2 \end{array} \right)
+
\left(\begin{array}{c}
   n_{11} \\
   n_{21} \\
   n^{*}_{12} \\
   n^{*}_{22} \end{array}\right),
\end{equation}
or more compactly
\begin{eqnarray}
\uy &=& s_1\ua+s_2\ub +\un, \text{ with} \nonumber \\
\uy &=& (y_{11},y_{21},y^{*}_{12},y^{*}_{22})^T,\nonumber \\
\ua &=& (h_{11},h_{21},h^{*}_{12},h^{*}_{22})^T, \nonumber \\
\ub &=& (h_{12},h_{22},-h^{*}_{11},-h^{*}_{21})^T,\text{ and }\nonumber \\
\un &=& (n_{11},n_{21},n^{*}_{12},n^{*}_{22})^T.
\end{eqnarray}
Since $\ua$ and $\ub$ are orthogonal the symbol estimates $\hat{s_1}$ and $\hat{s_2}$ can be determined by simply slicing $(\ua^{\dagger}\uy) / (\ua^{\dagger}\ua)$  and $(\uy^{\dagger}\ub)/(\ub^{\dagger}\ub)$ respectively.

Another advantage of the Alamouti method is that the densities of $\ua^{\dagger}\ua$ and $\ub^{\dagger}\ub$ are (identical and) chi-square with 8 degrees of freedom. This results in a diversity order 4, i.e.
\begin{equation}
\Pr\{\widehat{(S_1,S_2)} \neq (S_1,S_2) \} \approx \gamma'' \cdot \SNR^{-4},
\end{equation}
for fixed rate and large enough $\SNR$.

A disadvantage of the Alamouti method is that only two complex symbols are transmitted every two transmissions, but more-importantly that the symbols transmitted in the second transmission are more or less {\em repetitions} of the symbols in the first transmission.
Section \ref{sec:scarepsiso} however suggests that we can improve upon ordinary repetition.
\section{The rotated and scaled Alamouti method}
\subsection{Method description}
Having seen in section \ref{sec:scarepsiso} that scaled-repetition improves upon ordinary repetition in the SISO case, we use this concept to improve upon the standard Alamouti scheme for MIMO transmission.
Instead of just repeating the symbols in the second transmission we scale them.
More precisely, when $s_1$ and $s_2$ are elements of $\QAMz \define \{ a+jb | a \in \PAMv, b \in \PAMv \}$, we could transmit for some value of $\theta$ the signals
\begin{eqnarray} \label{eq:rotscalAlam}
\lefteqn{\left(\begin{array}{cc} x_{11} & x_{12} \\
                         x_{21} & x_{22} \end{array}\right) =
\left(\begin{array}{cc} s_{1}\cdot\exp(j\theta)  & -s^{*}_{2} \\
                        M_2(s_{2}) & M_2(s^{*}_{1}) \end{array}\right) }\nonumber \\
&=&
\left(\begin{array}{cc} s_{1}\cdot \exp(j\theta) & -s^{*}_{2} \\
                        2s_{2} & 2s^{*}_{1} \end{array}\right)
                        -
\left(\begin{array}{cc} 0               & 0 \\
                        D_2(s_{2}) & D_2(s^{*}_{1}) \end{array}\right),
\end{eqnarray}
where $M_2(\alpha) = 2\alpha - D_2(\alpha)$ with $D_2(\alpha) = 5\beta$ when $\beta$ is the complex sign of $\alpha$.

\begin{figure}
\centering
\epsfig{file=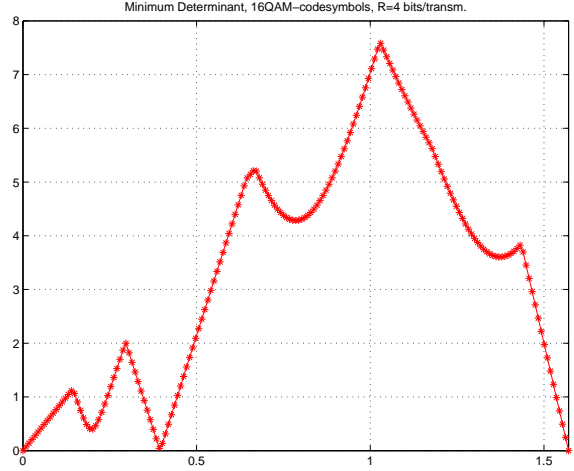,width=75mm}
\caption{Minimum modulus of the determinant for rotated and scaled Alamouti as a function of $\theta$ horizontally.}
\label{plot:detfortheta}
\end{figure}

A first question is to determine a good value for $\theta$.
Therefore we determine for $0 \leq \theta \leq \pi/2$ the minimum modulus of the determinant $\text{mindet}(\theta)$
\begin{equation}
\text{mindet}(\theta) = \min_{(s_1,s_2),(s'_1,s'_2)} |\det( X(s_1,s_2,\theta)- X(s'_1,s'_2,\theta))|,
\end{equation}
where $X =
\left(\begin{array}{cc} x_{11} & x_{12} \\
                        x_{21} & x_{22}  \end{array}\right)$ is the code matrix.
The minimum modulus of the determinant as a function of $\theta$ can be found in Fig. \ref{plot:detfortheta}.
The maximum value of the minimum determinant (i.e. 7.613) occurs for
\begin{equation}
\theta_{\text{opt.}}=1.028.
\end{equation}
We will use this value for $\theta$ in what follows.
\subsection{Hard-decision Performance}
\begin{figure}
\centering
\epsfig{file=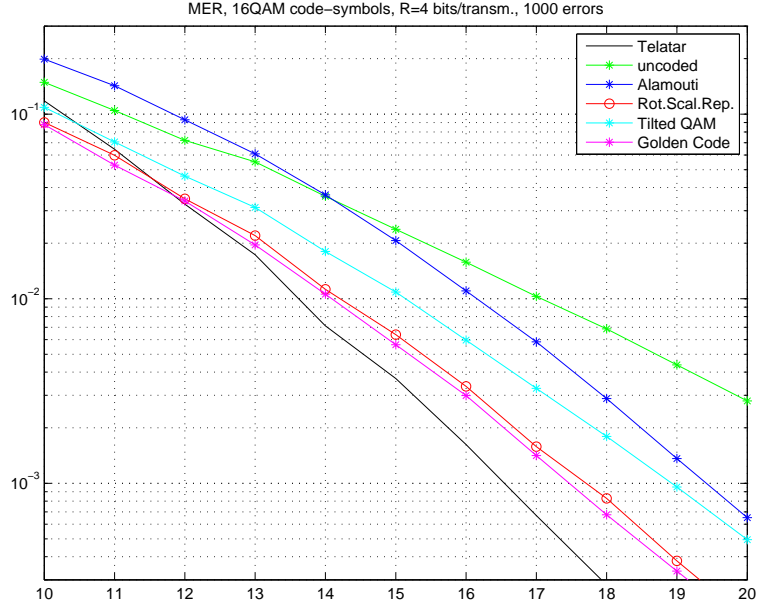,width=100mm}
\caption{Message error rate for several R=4 space-time codes.}
\label{plot:MER16}
\end{figure}

We have compared the message-error-rate for several $R=4$ space-time codes in Fig. \ref{plot:MER16}.
By message-error-rate we mean the probability $\Pr\{\widehat{X} \neq X \}$.
Note that for each "test" we generate a new message (8-bit) and a new channel matrix.
The decoder is optimal for all codes, it performs $ML$-decoding (exhaustive search).
The methods that we have considered are:
\begin{enumerate}
\item
{\bf Uncoded}, in green. We transmit
\begin{eqnarray}
X =
\left(\begin{array}{cc} x_{11} & x_{12} \\
                        x_{21} & x_{22}  \end{array}\right),
\end{eqnarray}
where $x_{11}, x_{12}, x_{21},$ and $x_{22}$ are symbols from $\QAMv$.
\item
{\bf Alamouti}, in blue, see  (\ref{eq:Alamouti}), where $s_1$ and $s_2$ are symbols from $\QAMz$.
\item
{\bf Tilted QAM}, in cyan. Proposed by Yao and Wornell \cite{YaoWorn2003}.
Let $s_a, s_b, s_c,$ and $s_d$ symbols from $\QAMv$. Then we transmit
\begin{eqnarray}
\left(\begin{array}{c} x_{11} \\ x_{22} \end{array} \right)
&=& \left(\begin{array}{cc} \cos(\theta_1) & -\sin(\theta_1) \\
                             \sin(\theta_1) & \cos(\theta_1) \end{array} \right) \left(\begin{array}{c} s_a \\ s_b \end{array} \right), \nonumber \\
\left(\begin{array}{c} x_{21} \\ x_{12} \end{array} \right)
&=& \left(\begin{array}{cc} \cos(\theta_2) & -\sin(\theta_2) \\
                             \sin(\theta_2) & \cos(\theta_2) \end{array} \right) \left(\begin{array}{c} s_c \\ s_d \end{array} \right),
\end{eqnarray}
for $\theta_1 = \frac12\arctan(\frac12)$ and $\theta_2 = \frac12\arctan(2)$.
\item
{\bf Rotated and scaled Alamouti}, in red, see (\ref{eq:rotscalAlam}) for $\theta=1.028$, and with $s_1$ and $s_2$ from $\QAMz$.
\item
{\bf Golden code}, in magenta. Proposed by Belfiore et al. \cite{BelfRekaVite2005}.
Now
\begin{eqnarray}
X = \
\frac{1}{\sqrt{5}} \left(\begin{array}{cc} \alpha(z_1+z_2\theta) & \alpha(z_3+z_4\theta) \\
                        j\cdot \overline{\alpha}(z_3+z_4\overline{\theta}) & \overline{\alpha}(z_1+z_2\overline{\theta})  \end{array}\right),
\end{eqnarray}
with $\theta=\frac{1+\sqrt{5}}{2}$, $\overline{\theta} = \frac{1-\sqrt{5}}{2}$, $\alpha=1+j-j\theta$, and  $\overline{\alpha}=1+j-j\overline{\theta}$ and where $z_1, z_2, z_3,$ and $z_4$ are $\QAMv$-symbols.
\item
{\bf Telatar}, in black. This is the probability that the Telatar capacity of the channel is smaller than 4.
\end{enumerate}
Clearly it follows from Fig. \ref{plot:MER16} that the winner is the Golden code. However rotated and scaled Alamouti is only slightly worse, roughly $0.2$ dB.
Important is that Alamouti coding is roughly 2 dB worse than the Golden code.

\section{Decoding complexity}
Clearly the Golden code is better than rotated and scaled Alamouti.
However the Golden code in principle requires the decoder to check all 256 alternative codewords.
Here we will investigate the complexity and performance of a {\bf suboptimal} rotated and scaled Alamouti decoder.
Denote $\Theta=\exp(j\theta_{\text{opt.}})$.

A. In the rotated and scaled Alamouti case the received vector is
\begin{eqnarray}
\lefteqn{
\left(\begin{array}{c}
   y_{11} \\
   y_{21} \\
   y^{*}_{12}\\
   y^{*}_{22} \end{array}\right)
=
\left(\begin{array}{cc}
        h_{11} \Theta        & 2h_{12}  \\
        h_{21} \Theta       & 2h_{22} \\
        2h^{*}_{12} & -h^{*}_{11} \\
        2h^{*}_{22} & -h^{*}_{21} \\
      \end{array} \right)
\left(\begin{array}{c} s_1 \\ s_2 \end{array} \right) } \\
&& -
\left(\begin{array}{c}
   0 \\
   0 \\
   h^{*}_{12} \\
   h^{*}_{22} \end{array}\right) D_2(s_1)
- \left(\begin{array}{c}
   h_{12} \\
   h_{22} \\
   0 \\
   0 \end{array}\right) D_2(s_2)
+ \left(\begin{array}{c}
   n_{11} \\
   n_{21} \\
   n^{*}_{12} \\
   n^{*}_{22} \end{array}\right). \nonumber
\end{eqnarray}
We can write this as
\begin{eqnarray*}
\uy &=& s_1\ua+s_2\ub - D_2(s_1) \uc - D_2(s_2) \ud +\un, \\
\uy &=& (y_{11},y_{21},y^{*}_{12},y^{*}_{22})^T,\\
\ua &=& (h_{11}\Theta,h_{21}\Theta,2h^{*}_{12},2h^{*}_{22})^T, \\
\ub &=& (2h_{12},2h_{22},-h^{*}_{11},-h^{*}_{21})^T,\\
\uc &=& (0,0,h^{*}_{12},h^{*}_{22})^T,\\
\ud &=& (h_{12},h_{22},0,0)^T, \text{ and }\\
\un &=& (n_{11},n_{21},n^{*}_{12},n^{*}_{22})^T.
\end{eqnarray*}
For the $\cos(\phi)$ of the angle between $\ua$ and $\ub$ we can write
\begin{equation}
\cos (\phi) = \frac{|  2(\Theta-1)(h_{11}h_{12}^{*}+h_{21}h_{22}^{*})] }{ |h_{11}|^2 + |h_{21}|^2 + 4|h_{12}|^2 + 4|h_{22}|^2}.
\end{equation}
B. Instead of decoding $(s_1,s_2)$ we can also decode $(t_1,t_2) = (M_2(s_1),M_2(s_2))$ which is equivalent to $(s_1,s_2)$.
Therefore we rewrite (\ref{eq:rotscalAlam}) and obtain
\begin{eqnarray}
\lefteqn{\left(\begin{array}{cc} x_{11} & x_{12} \\
                         x_{21} & x_{22} \end{array}\right) =
\left(\begin{array}{cc} -M_2(t_{1})\Theta  & M_2(t^{*}_{2}) \\
                        t_{2} & t^{*}_{1} \end{array}\right) }\nonumber \\
&=&
\left(\begin{array}{cc} -2t_{1}\Theta & 2t^{*}_{2} \\
                        t_{2} & t^{*}_{1} \end{array}\right)
                        -
\left(\begin{array}{cc}  -D_2(t_{1})\Theta & D_2(t^{*}_{2})\\
                         0               & 0
                        \end{array}\right),
\end{eqnarray}
since $t=M_2(s)$ implies that $s=-M_2(t)$.
Now
\begin{eqnarray}
\lefteqn{
\left(\begin{array}{c}
   y_{11} \\
   y_{21} \\
   y^{*}_{12}\\
   y^{*}_{22} \end{array}\right)
=
\left(\begin{array}{cc}
        -2 h_{11} \Theta & h_{12}  \\
        -2 h_{21} \Theta & h_{22} \\
        h^{*}_{12}      & 2 h^{*}_{11} \\
        h^{*}_{22}      & 2 h^{*}_{21} \\
      \end{array} \right)
\left(\begin{array}{c} t_1 \\ t_2 \end{array} \right) } \\
&& -
\left(\begin{array}{c}
   -h_{11} \Theta \\
   -h_{21} \Theta \\
   0 \\
   0 \end{array}\right) D_2(t_1)
- \left(\begin{array}{c}
   0 \\
   0 \\
   h^{*}_{11} \\
   h^{*}_{21}
   \end{array}\right) D_2(t_2)
+ \left(\begin{array}{c}
   n_{11} \\
   n_{21} \\
   n^{*}_{12} \\
   n^{*}_{22} \end{array}\right). \nonumber
\end{eqnarray}
We can write this as
\begin{eqnarray*}
\uy &=& t_1\ua'+t_2\ub' - D_2(t_1) \uc' - D_2(t_2) \ud' +\un, \\
\ua' &=& (-2h_{11}\Theta,-2h_{21}\Theta,h^{*}_{12},h^{*}_{22})^T, \\
\ub' &=& (h_{12},h_{22},2h^{*}_{11},2h^{*}_{21})^T,\\
\uc' &=& (-h_{11}\Theta,-h_{21}\Theta,0,0)^T,\text{ and } \\
\ud' &=& (0,0,h^{*}_{11},h^{*}_{21},0,0)^T, \\
\end{eqnarray*}
and for the $\cos(\phi')$ of the angle between $\ua'$ and $\ub'$ we can write
\begin{equation}
\cos(\phi') = \frac{|  2(\Theta-1)(h_{11}h_{12}^{*}+h_{21}h_{22}^{*})| }{ 4|h_{11}|^2 + 4|h_{21}|^2 + |h_{12}|^2 + |h_{22}|^2}.
\end{equation}
C. It now follows from the inequality $2r_1r_2 \leq r_1^2 + r_2^2$ (where $r_1$ and $r_2$ are reals), that
\begin{eqnarray}
\cos (\phi)  &\leq & |\Theta-1| \cdot
\frac{ |h_{11}|^2 +|h_{12}|^2 + |h_{21}|^2 +|h_{22}|^2 }
     {  |h_{11}|^2 + |h_{21}|^2 + 4|h_{12}|^2 +4|h_{22}|^2}, \nonumber \\
\cos (\phi') &\leq & |\Theta-1| \cdot
\frac{ |h_{11}|^2 +|h_{12}|^2 + |h_{21}|^2 +|h_{22}|^2 }
     { 4|h_{11}|^2 +4|h_{21}|^2 +  |h_{12}|^2 + |h_{22}|^2}.
\end{eqnarray}
If
\begin{equation} \label{condition}
|h_{12}|^2+ |h_{22}|^2 \geq |h_{11}|^2 +|h_{21}|^2,
\end{equation}
then $\cos(\phi)  \leq \frac{2|\Theta-1|}{5} = 0.393$, else $\cos(\phi')  \leq \frac{2|\Theta-1|}{5} = 0.393$.
Therefore it makes sense to decode $(s_1,s_2)$ when (\ref{condition}) holds and $(t_1,t_2)$ when (\ref{condition}) does not hold.
Using zero-forcing to decode, the noise enhancement is then at most $1/(1-0.393^2) = 1.183$ which is 0.729 dB.
We shall see later that noise enhancement turns out to be un-noticeable in practise.

D. The decoding procedure is straightforward. Focus on the case where we decode $(s_1,s_2)$ for a moment. For all 16 alternatives of $(D_2(s_1),D_2(s_2))$ the vector
\begin{equation}
\uz = \uy + D_2(s_1)\uc + D_2(s_2)\ud =  s_1 \ua + s_2 \ub + \un
\end{equation} and
is determined.
Then compute the sufficient statistic
\begin{equation}
\left( \begin{array}{c}
   \ua^{\dagger} \uz \\
   \ub^{\dagger} \uz  \end{array}\right)
   = \left( \begin{array}{cc}
   \ua^{\dagger} \ua & \ua^{\dagger}\ub   \\
   \ub^{\dagger} \ua & \ub^{\dagger}\ub  \end{array}\right)
   \left( \begin{array}{c}
   s_1 \\
   s_2 \end{array}\right) +
\left( \begin{array}{c}
   \ua^{\dagger} \un \\
   \ub^{\dagger} \un  \end{array}\right).
\end{equation}
Next use the inverted matrix $M =
\left( \begin{array}{cc}
   \ub^{\dagger} \ub & -\ua^{\dagger}\ub   \\
   -\ub^{\dagger} \ua & \ua^{\dagger}\ua  \end{array}\right) / D$ where $D= (\ua^{\dagger}\ua)(\ub^{\dagger}\ub) - (\ub^{\dagger}\ua)(\ua^{\dagger} \ub)$ to obtain
$\left( \begin{array}{c}
   \tilde{s_1}  \\
   \tilde{s_2}  \end{array}\right) = M  \left( \begin{array}{c}
   \ua^{\dagger} \uz \\
   \ub^{\dagger} \uz  \end{array}\right)$.
Next both $\tilde{s_1}$ and $\tilde{s_2}$ are sliced under the restriction that only alternatives that match the assumed values $D_2(s_1)$ and $D_2(s_2)$ are possible outcomes.
This is done for all 16 alternatives $(D_2(s_1),D_2(s_2))$.
The best result in terms of Euclidean distance is now chosen.

In considering all alternatives $(D_2(s_1),D_2(s_2))$ we only need to slice when the length of $\uz - \tilde{s_1}\ua - \tilde{s_2}\ub$ is smaller than the closest distance we have observed so far.
This reduces the number of slicing steps. We call this approach METHOD 1.

E. The number of slicing steps can even be further decreased if we start slicing with the most promising alternative  $(D_2(s_1),D_2(s_2))$. This approach is called METHOD 2.
Therefore we note that the "direct" $s_1$-signal-component in $X$ is $\left(\begin{array}{cc}
        s_{1} \Theta & 0  \\
        0  & - s_1^*/2  \\
      \end{array} \right)$.
Therefore we can slice $(\ue_1^{\dagger} \uy)/ (\ue_1^{\dagger} \ue_1)$ in order to find a good guess for $D_2(s_1)$.
Similarly we slice $(\ue_2^{\dagger} \uy)/ (\ue_2^{\dagger} \ue_2)$ to find a good first guess for $D_2(s_2)$.
Here
\begin{eqnarray}
\ue_1 &=& (h_{11}\Theta,h_{21}\Theta,-h^{*}_{12}/2,-h^{*}_{22}/2)^T, \\
\ue_2 &=& (-h_{12}/2,-h_{22}/2,-h^{*}_{11},-h^{*}_{21})^T.
\end{eqnarray}
Then we consider the other 15 alternatives and only slice if necessary.
Note that similar methods apply if we want to decode $(t_1,t_2)$.

\begin{figure}
\centering
\epsfig{file=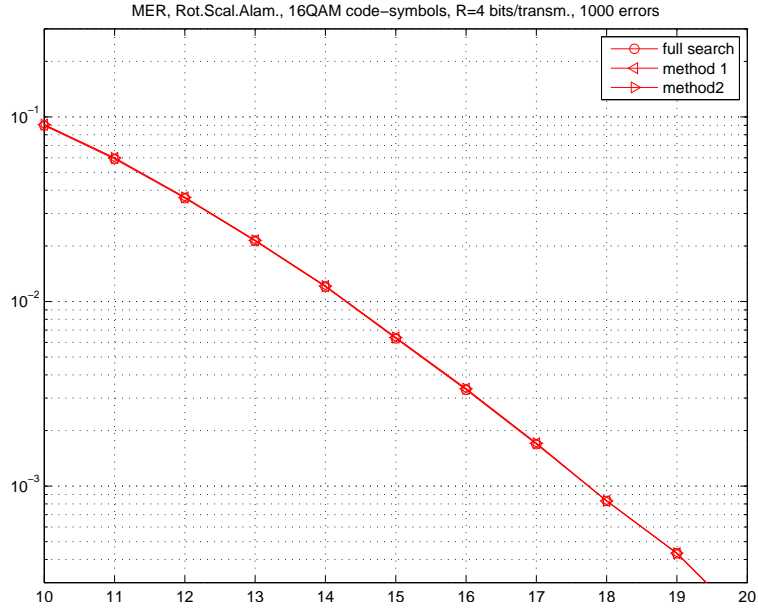,width=100mm}
\caption{Message error rate for three Rotated Scaled Alamouti decoders ($R=4$), horizontally $\SNR$.} \label{plot:MER16RSA}
\end{figure}

\begin{figure}
\centering
\epsfig{file=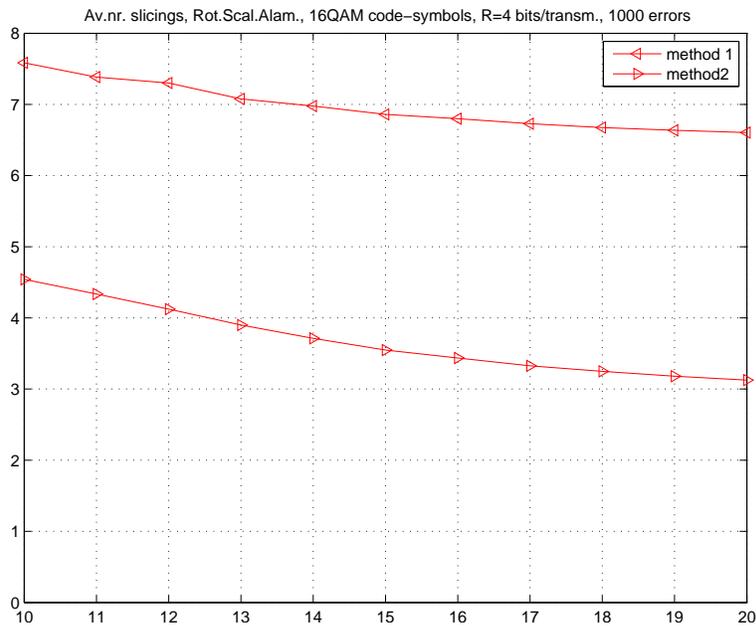,width=100mm}
\caption{Number of slicings for two Rotated Sclaed Alamouti decoders ($R=4$), horizontally $\SNR$.}
\label{plot:SLICINGS16}
\end{figure}

F. We have carried out simulations, first to find out what the degradation of the suboptimal decoders according to method 1 and method2 is relative to ML-decoding.
The result is shown in Fig. \ref{plot:MER16RSA}.
Conclusion is that the suboptimal decoders do not demonstrate a performance degradation.
We have also considered the number of slicings for both method 1 and method 2.
This is shown in Fig. \ref{plot:SLICINGS16}.
It can be observed that method 1 leads to roughly 7 slicings (as opposed to 16).
Method 1 further decreases the number of slicing to roughly 3.5.

\section{Conclusion}
Rotated and scaled Alamouti has a hard-decision performance which is only slightly worse than that of the Golden code, but can be decoded with an acceptable complexity.
We finally remark that we have obtained similar results for codes based on mapping $M_3(\cdot)$ for $9$-PAM.

\end{document}